\documentclass[letterpaper]{article}
\usepackage{aaai}
\usepackage{graphicx}
\usepackage{tabularx}
\usepackage{amsmath}
\usepackage{float}
\usepackage{multirow}
\usepackage{xcolor}
\newcommand{\malrnn}{M\lowercase{a}lRNN }
\usepackage{times}
\usepackage{helvet}
\usepackage{courier}
\usepackage{soul}
\begin{document}
%
\title{Binary Black-box Evasion Attacks Against Deep Learning-based Static Malware Detectors with Adversarial Byte-Level Language Model}

\author{Mohammadreza Ebrahimi,\textsuperscript{1}
Ning Zhang,\textsuperscript{2}
James Hu,\textsuperscript{1}
Muhammad Taqi Raza,\textsuperscript{1}
Hsinchun Chen\textsuperscript{1}\\
\textsuperscript{1}{Artificial Intelligence Lab, The University of Arizona}\\
\textsuperscript{2}{Covax Data Inc., Arizona}\\
\{ebrahimi, jameshu, taqi, hsinchun\}@email.arizona.edu; ning.zhang@covaxdata.com
}

\maketitle
\begin{abstract}
\begin{quote}
Anti-malware engines are the first line of defense against malicious software. While widely used, feature engineering-based anti-malware engines are vulnerable to unseen (zero-day) attacks. Recently, deep learning-based static anti-malware detectors have achieved success in identifying unseen attacks without requiring feature engineering and dynamic analysis. However, these detectors are susceptible to malware variants with slight perturbations, known as adversarial examples. Generating effective adversarial examples is useful to reveal the vulnerabilities of such systems. Current methods for launching such attacks require accessing either the specifications of the targeted anti-malware model, the confidence score of the anti-malware response, or dynamic malware analysis, which are either unrealistic or expensive. We propose MalRNN, a novel deep learning-based approach to automatically generate evasive malware variants without any of these restrictions. Our approach features an adversarial example generation process, which learns a language model via a generative sequence-to-sequence recurrent neural network to augment malware binaries. MalRNN effectively evades three recent deep learning-based malware detectors and outperforms current benchmark methods. Findings from applying our MalRNN on a real dataset with eight malware categories are discussed. 
\end{quote}
\end{abstract}

\section{Introduction}
\label{introduction}
\noindent Malware attacks pose a massive threat to the security of companies and individuals. The average annual cost of malware attacks has increased to \$2.6 million per mid-sized company worldwide \cite{bissell2019}. Anti-malware engines are essential to proactively prevent these attacks  \cite{tounsi2018survey}. Most anti-malware engines mainly rely on signature-based approaches that match manually-defined patterns against known malicious files \cite{anderson2018learning}. The success of signature-based methods significantly depends on the quality and recency of the pre-defined rules that are often handcrafted by malware analysts. While useful, signature-based engines suffer from two significant deficiencies: first, they could be ineffective in dealing with newly evolved variants of malware, and thus, vulnerable to ‘unseen’ variants known as zero days \cite{chenB2019adversarial}; second, they rely on manually defined rules that cannot keep up with the rapid evolution of malware variants.
    
Due to the deficiencies of signature-based anti-malware engines, researchers have  presented machine learning-based malware detection. However classic machine learning algorithms often require manual feature engineering. Recently, a new stream of Deep Learning (DL)-based malware detector has emerged that can consume the whole raw malware binary as input and extract the salient features automatically, without relying on manually defined rules or feature engineering. As a result, successful DL-based anti-malware engines have emerged \cite{raff2018malware}, \cite{fleshman2019non}, \cite{krvcal2018deep}. However, DL-based anti-malware engines have shown to be susceptible to small perturbations in their input, featured by automated attacks known as Adversarial Example Generation (AEG) \cite{demetrio2019explaining}. These attacks yield slightly perturbed malware variants that can mislead the DL-based engines into miss-classifying them as benign. Given the crucial role of anti-malware in preventing cyber-attacks and improving the security posture of many organizations, there is a vital need to devise automatic ways to protect anti-malware engines against the AEG attacks.

Although AEG can negatively affect the performance of DL-based engines, it can also be utilized to further improve their performance. Anti-Malware Evasion (AME) has emerged as a promising method to automate the AEG process for this purpose \cite{chenY2019training}. Malware variants that successfully evade the DL-based malware detectors can be employed in re-training and improving them. Moreover, verifying DL-based anti-malware engines against AEG is a viable defense mechanism \cite{goodfellow2018making}. In effect, automatic emulation of AEG attacks can help strengthen the ability of DL-based engines to detect malware.
    
AME methods often rely on additive approaches, which inject bytes into the malware binary, known as append attacks \cite{suciu2019exploring}. Append attacks are a natural fit for AME because they do not affect the functionality of the malware since their injected payload is not executed by the operating system and thus they do not interfere with the malware execution \cite{castroandbiggio2019poster}, \cite{suciu2019exploring}. Nevertheless, current approaches for launching these attacks suffer from two major issues that limit their applicability. First, many attacks assume full knowledge about the anti-malware architecture, its parameters, or the confidence level of the anti-malware response. These assumptions do not apply to realistic attack scenarios in which the information is hidden from the adversary \cite{hu2018black}. Second, since they often rely on brute-force mechanisms to craft new malware variants, they require a high volume of appended bytes (i.e., payload) to evade the anti-malware engine \cite{suciu2019exploring}.

Deep learning methods have shown promise in generating smaller and more effective perturbations \cite{kreuk2018adversarial}. Recently, among deep learning methods, deep language models have shown promise in malware analysis by treating the malware binary sequence as characters in a written language \cite{awad2018modeling}. Generative Recurrent Neural Network (RNN) is a powerful architecture to learn such language models \cite{mogren2019character}. Motivated by the importance of finding the vulnerabilities of current DL-based anti-malware engines, we propose a new threat model that utilizes a novel RNN-based method to automatically construct adversaries for evading several DL-based anti-malware engines simultaneously. To this end, we focus on how to automatically generate evasive malware samples on a large scale. Our study offers a novel approach to directly learn a language model on binary executables and generate benign-looking content without requiring \textit{any} knowledge of the targeted anti-malware. To our knowledge, the proposed method contributes to the first automated attack against DL-based anti-malware engines without these restrictive assumptions. Furthermore, our approach does not require expensive dynamic malware analysis. To foster reproducibility, we made the code and the dataset available to the AI-enabled security research community on GitHub at {\color{blue}\underline{https://github.com/johnnyzn/MalRNN}}.

\section{Background and Related Work}
\label{background}

\subsection{Adversarial Example Generation (AEG)}

Deep learning models have been recently shown to fail when an adversary carefully modifies their input data with subtle perturbations. Adversarial examples are instances with meticulous feature perturbations that can cause a target machine learning model to make wrong decisions. Automatically crafting such instances by an adversary against a specific class of target machine learning models is an emerging task in artificial intelligence, referred to as AEG  \cite{goodfellow2018making}. This concept of AEG that we use in this study is not meant to be confused with Automatic Exploit Generation). Verifying machine learning models against AEG is a crucial defense mechanism that not only helps improve the resistance of these models, but also provides insights for designing better machine learning models \cite {goodfellow2018making}.

Depending on the information available to the adversary from the targeted machine learning model, AEG is carried out under four possible scenarios \cite{qiu2019review,anderson2018learning}. In the first scenario, known as a white-box attack, the adversary has full access to the structure and parameters of the attack target. The second AEG scenario is referred to as gray-box AEG and pertains to situations in which the parameters of the attacked neural network model are not available but the adversary has access to the features that are important for decision making by target classifier. The third scenario, called black-box AEG, relates to when the adversary cannot access the model's specification, features, or parameters; however, it can obtain a real-valued feedback, also known as confidence score, from the attack target. Finally, binary black-box AEG applies to a black-box scenario in which not only no a priori knowledge is assumed about the target, but also the adversary does not have access to a real-valued feedback from the attack target. Instead, in binary black-box scenario, the adversary can only observe a binary response associated with the success or failure of the crafted instance in evading the attack target. This type of attack is also known as binary black box \cite{anderson2018learning}. Binary black-box AEG is the most restrictive and the most common scenario in real-world \cite{fleshman2019non}, since oftentimes the specification and confidence score of the attack target are unknown.

\subsection{Anti-Malware Evasion}
    Conducting AEG in the Malware detection domain gives rise to anti-malware evasion (AME) attacks, a new stream of research that employs AEG to perturb malware samples and generate variants that evade anti-malware engines while still preserving the functionality of the original malware. AME attacks can be categorized based on the type of threat model they implement (i.e., white, gray, black, and binary black-box). Consistent with our goal of proposing a more realistic AME attack scenario in our study, we examine the past AME studies that support black-box and binary black-box attacks.
    Among these studies, AME studies that offer black-box attacks do not require knowing the specifications of the targeted anti-malware \cite{demetrio2020efficient,castroandbiggio2019poster,castroandschmitt2019armed,chenB2019adversarial,park2019generation,suciu2019exploring,hu2018black}. These studies employ a wide range of methods such as genetic algorithm \cite{demetrio2020efficient}, random perturbations \cite{castroandschmitt2019armed,chenB2019adversarial}, dynamic programming \cite{park2019generation}, and RNN \cite{hu2018black}. However, these methods heavily rely on the confidence score feedback obtained from anti-malware engines to craft their perturbations. The confidence score is a real value ranging between 0 and 1, which indicates the probability that the input is malware. This value is interpreted as the confidence of the decision made by the anti-malware engine. While black-box attacks are more realistic than white-box attacks, the confidence score is \textit{internal} to the anti-malware engine and thus not visible to the adversary. This issue restricts the usability of these methods \cite{rosenberg2019defense}.
    
    Binary black-box attacks, on the other hand, do not require observing the anti-malware's confidence score \cite{dey2019evadepdf,fang2019evading,rosenberg2019defense,anderson2018learning}, and thus, are applicable to real attack scenarios. Nevertheless, most current binary black-box AME studies target signature-based anti-malware engines \cite{dey2019evadepdf,fang2019evading,anderson2018learning}. Although Rosenberg et al. \cite{rosenberg2019defense} propose a binary black-box attack on DL-based anti-malware engines, their approach requires an API call sequence obtained from expensive and time-consuming dynamic analysis of malware binary in a sandbox. We also note that Hu and Tan \cite{hu2018black} propose a black-box AME attack that is based on training an RNN. However, similar to \cite{rosenberg2019defense}, their approach requires a sequence of API calls obtained during the dynamic analysis in a sandbox. Another practical limitation of the current binary black-box AME methods relates to the brute-force operationalization of append attacks, which requires a large size of binary content to be appended to the original malware file (e.g., three times larger than the size of the original malware) \cite{suciu2019exploring,castroandschmitt2019armed}. This results in generating abnormally large malware variants that can be detected by anti-malware engines due to the suspicious size of the resulting malware variant. Furthermore, most AME studies on attacking DL-based anti-malware engines are designed to only target a specific anti-malware architecture with certain parameter settings. Focusing on evading one specific architecture limits the generalizability of such methods to other anti-malware models. We expect that learning a universal language model from benign executables can facilitate attaining more generalizable AME methods.

\subsection{Generative RNN-based Language Models}
\label{background_RNN}
Constructing a language model amounts to learning a probability distribution over a sequence of strings or characters. Once learned, a language model can be used to \textit{generate} the next element in a given sequence. Neural language models with recurrent architectures have shown promise in generating high-quality sequences in Natural Language Processing (NLP) tasks \cite{kim2016character}. A generative RNN processes sequential input while preserving temporal patterns in the sequence. At each time step $t$, an RNN takes an input $x_t$ and the current hidden state $h_t$ to emit a continuous value. This value is used to generate/predict future elements in the sequence. This generative nature of RNNs makes them suitable for sequence analysis tasks such as language modeling \cite{belletti2019quantifying}, where the elements of the input are time-dependent. Once trained, RNNs yield effective language models on short natural language text and binary content \cite{zuo2018neural} that are able to predict the next element based on a given input sequence.
Two major challenges arise in utilizing RNN-based language models on malware content. First, using RNN language models for learning long sequences of malware content is challenging \cite{raff2018malware} due to the large number of time steps, which leads to the attenuation of the error signal during training, widely known as vanishing gradient problem \cite{goldberg_neural_2017}. Adding gating mechanism to the input and output of RNN units can address this issue and yields an effective variant of RNN, Gated Recurrent Units (GRU) \cite{goldberg_neural_2017}.
Generative RNNs require the input and output sequences to have the same dimensions. While this is useful in machine learning tasks such as part of speech tagging, it limits their applicability in the malware domain. Among RNN-based architectures for language modeling, \emph{sequence-to-sequence} models address this issue by adding an additional encoding step before feeding the data to the generative RNN. Sequence-to-sequence models have recently yielded breakthrough results in many sequence analysis tasks such as machine translation \cite{ono-etal-2019-hybrid} and speech recognition \cite{irie2019choice}. They can map the input sequence of a fixed length to a generated output sequence of a different length. Given their recent success in other machine learning fields, we expect that sequence-to-sequence RNN-based language models can provide an effective tool to automatically generate benign-looking adversarial examples for AME applications. Accordingly, we propose to construct an RNN language model directly on the binary content (as opposed to the sequence of API calls in a sandbox) to accomplish adversarial malware generation in a binary black-box scenario without requiring dynamic analysis.    
    
\section{Proposed Method (\malrnn)}
\label{method}

As noted in Section 2, most black-box AME methods rely on a brute-force approach in which they inject bytes into a malware sample until the generated variant evades the anti-malware. The brute-force property of these methods leads to crafting variants with large payload size that renders AME less effective. This issue motivates a threat model that limits the volume of injected bytes, as opposed to the one that allows adding an indefinite length of perturbations.

\begin{figure*}[t]
    \centering
    \includegraphics[width=1\textwidth]{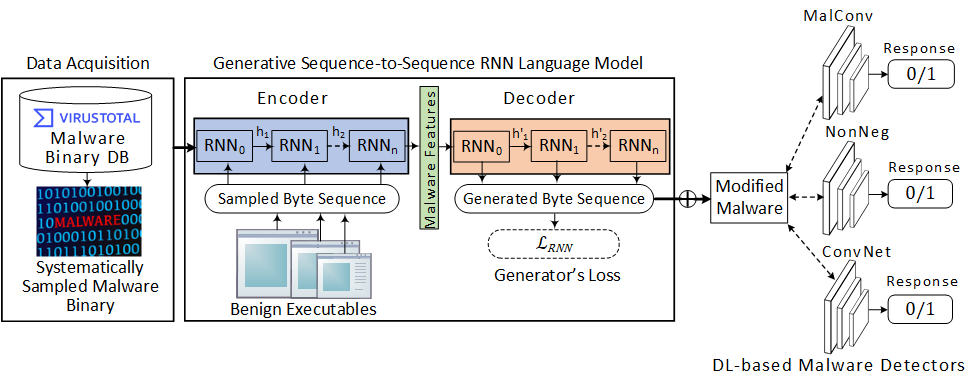}
    \caption{Abstract view of \malrnn malware evasion architecture}
    \label{architecture}
\end{figure*}

\subsection{Threat Model}
Consistent with \cite{anderson2018learning}, we define the threat model for launching binary black-box AME attacks against static anti-malware models. Nevertheless, unlike the threat model proposed in \cite{anderson2018learning}, which targets feature-based anti-malware engines, our threat model focuses on launching attacks against DL-based anti-malware engines. Three major components of our threat model are:
    
    \begin{itemize}
        \item{\verb|Adversary’s Goal:|} Automatically crafting malware variants that are capable of evading DL-based anti-malware.
        \item{\verb|Adversary’s Knowledge:|} The structure and parameters of the anti-malware model are unknown to the adversary. Furthermore, the adversary does not have access to the confidence score produced by anti-malware. The only information available to the adversary is whether the generated malware variant can evade the anti-malware or not.
        \item{\verb|Adversary's Capability:|} Applying functionality preserving append modifications on malware binary, while the maximum modification size is limited. We focus on append modifications, since they very often do not interfere with the functionality of the malware.
    \end{itemize}

To realize this threat model, we propose MalRNN, a byte-level sequence-to-sequence generative model that learns a language model on benign samples and injects benign-looking byte sequences into the original malware binary in order to obtain evasive malware variants.

\subsection{\malrnn Design}
\label{designing_malrnn}

In accordance with the above threat model, it can be expected that that mimicking the patterns of benign executables could be a viable attack approach. We incorporate this insight into our design of MalRNN. Specifically, this is achieved through learning a language model on bytes that can generate benign-looking samples. Such a language model significantly contributes to alleviating brute-force trial and error for generating evasive variants. Figure \ref{architecture} illustrates the major components of our MalRNN malware evasion architecture. We describe each component in the remaining of this section.

\subsection{Data Acquisition}
Developing \malrnn requires two datasets of binary executables: 1) a malware executable dataset that serves as the initial seed to generate evasive malware variants, and 2) a benign executable dataset to train the language model. To obtain the former dataset, we compiled an up-to-date collection with recent real malware samples from the last three years. The dataset includes over 6,000 malware binaries from eight common malware categories. The distribution of the dataset is described later. To obtain the benign executable dataset, following \cite{raff2018malware}, we collected 4,329 benign executables from a clean installation folder of Microsoft Windows. In both datasets, we converted the binary input to hexadecimal characters suitable for processing by a character-level language model. Furthermore, to avoid inefficient training with long input byte sequences in malicious and benign executables, we employed systematic sampling. This process samples the input binary sequence in fixed intervals to reduce the input size for generative sequence-to-sequence RNN language model.

\subsection{Generative Sequence-to-Sequence RNN Language Model}
Our model employs a character-level sequence-to-sequence RNN to learn a language model from \emph{benign} malware binaries. We adopt Gated Recurrent Units (GRU) \cite{goldberg_neural_2017} as the building block of our sequence-to-sequence model to alleviate the gradient vanishing problem in processing long sequences,\cite{dey2017gate}. \malrnn aims to maximize the adversarial loss \cite{madry2017towards} of the anti-malware model, which is formulated in the following equation:

\begin{equation}
  \underset{\delta \in \Delta}{\operatorname{maximize}}\  \mathcal{L}(\mathcal{H}_\theta(x+\delta),y)
\end{equation}

where $x$ is the input malware sample, $\mathcal{H_\theta}$ is the attacked DL-based anti-malware model parameterized by $\theta$, and $\delta \in \Delta$ denotes the allowable perturbations that preserve functionality (appending byte sequences in our case). The loss function $\mathcal{L}$ represents binary cross-entropy loss in most DL-based anti-malware engines. However, in a (binary) black-box setting the exact loss function from the anti-malware is not accessible and thus, cannot be incorporated into the model's loss. Accordingly, directly maximization of Eq. 1 is impractical. The key idea behind our model is that maximizing the loss in Eq. 1 translates to minimizing the loss of an adversarial model in generating benign-looking samples that can bypass the anti-malware. The middle box in Figure \ref{architecture} shows our character-level generative RNN for learning such an adversarial model serving as a binary content generator.

Inspired by the recent sequence-to-sequence RNN in language modeling, \malrnn's generator consists of two main RNN components: An encoder RNN and a decoder RNN. The encoder aims to encapsulate the salient features of the input byte sequence into a feature vector. This vector is obtained from the final hidden state of the last RNN unit in the encoder architecture and is fed to the decoder RNN (shown in the vertical inner box in Figure \ref{architecture}). 

The encoder's current hidden states $h_t$ is obtained as a function of both its previous state $h_{t-1}$ and the current input element $x_t$. More formally, $h_t$ is given by Equation \ref{encoder_hidden}:
\begin{equation}
  h_t = f(W^{h}h_{t-1} + W^{x}x_t)
  \label {encoder_hidden}
\end{equation}
where $W^{h}$ denotes the network weights between the hidden units and and $W^{x}$ represents the network weights between hidden units and the input elements. Function $f$ is a nonlinear activation such as $tanh(.)$. The decoder receives the feature vector from the encoder and reconstructs the byte sequence that minimizes a cross-entropy loss between the generated bytes and benign samples ($\mathcal{L}_{RNN}$) at each time step. Unlike the encoder, the decoder's hidden state at each time step is only a function of the previous hidden state and is given by Equation \ref{decoder_hidden}:

\begin{equation}
  h_t = f(W^{h}h_{t-1})
  \label {decoder_hidden}
\end{equation}

 After training is complete, the generator learns to append benign-looking binary content to the malware binary in order to maximize the adversarial loss and construct an evasive malware variant. To craft a candidate malware variant, after completion of each training iteration, the generated byte sequence from MalRNN's generator is attached to the original malware. The candidate malware variant is checked against one or more black-box anti-malware models to assess if it can evade them. The output of the anti-malware is a binary output with 1 and 0 denoting detection and evasion, respectively. If the generated malware variant successfully evades the detectors, the candidate sample is saved as an evasive variant and will be further processed for ensuring its functionality. 

Such a model is suitable for launching binary black-box attacks described in our threat model since it depends neither on the gradients obtained from a differentiable anti-malware model, as in \cite{castroandbiggio2019poster,kolosnjaji2018adversarial}, nor on the confidence score received from the anti-malware engine, as in \cite{chen2017zoo}. This amounts to achieving an adversary that is agnostic to the targeted anti-malware's deep learning architecture. It is worth noting that, following \cite{anderson2018learning}, in order to comply with the binary black-box attack scenario, the confidence score provided by anti-malware architectures was masked to mimic a binary output from anti-malware.

In each iteration, \malrnn is trained on a sample of benign executables and generates a byte sequence that is appended to the end of the original anti-malware to form a new variant, which subsequently is tested against the targeted black-box anti-malware models. In each iteration, the RNN is trained on a sample of benign files, and generates the new bytes based on a given malware sequence. This process repeats until the new variant evades the anti-malware or the maximum number of attempts is reached. In case the maximum number of attempts for a specific input sample is reached the model proceeds to the next malware sample.

We implemented \malrnn using PyTorch. \malrnn was run on a single Nvidia RTX 2080 GPU with 4,352 CUDA cores and 8 GB internal memory. The code is designed to run on both GPU and CPU environments. The data comprises the full testbed including the benign executables for training the language model and also the malware binary dataset. MalrRNN's specifications, including the architecture and (hyper) parameter settings are given in Appendix A.

\subsection{Ensuring the Functionality of Generated Malware Variants} \label{ensure_func}
We used VirusTotal’s API, which supports large-scale malware analysis, for assessing the functionality of malware samples after modification. VirusTotal provides a malware behavior report that includes static and dynamic analysis of the malware sample. These reports describe network behavior, file access behavior, etc. Using the VirusTotal API, we compare the behavior reports for the modified evasive variants and original (i.e., unmodified) malware samples. Through this process, we ensure that the key parts of the Virus Total's report stay the same after modification, showing that the modified malware samples can be executed on the operating system and are fully functional. All 6,037 malware samples in our dataset were checked to be functional after appending bytes to their overlay. That is, the non-functional samples in the original dataset (more than 90\%) were excluded from the evaluation.

\section{Implementation and Evaluation}
\label{evaluation}
\subsection{Testbed and Evaluation Criterion}
\label{evaluation_criteria}
We obtained an academic license of VirusTotal and extracted 6,307 recent malware binaries from the past three years (2017-2019) in eight categories, including botnet, ransomware, spyware, adware, virus, dropper, backdoor, and rootkit. Table \ref{testbed} shows the distribution of the dataset by malware category. To be able to gain insight into each specific malware category, we evaluate \malrnn's performance on each category separately.
Utilizing the functionality assessment process described earlier, we checked the functionality of all modified malware binary samples to ensure they retain their functionality after modification. 

\begin{table}[ht]
\small
\caption{Breakdown of testbed based on different malware categories}
\begin{tabularx}{0.45\textwidth} { 
  | >{\centering\arraybackslash}X
  | >{\centering\arraybackslash}p{3cm}  
  | >{\centering\arraybackslash}X  | }
 \hline
 \textbf{Malware Category} & \textbf{Examples} & \textbf{\# of Malware Samples} \\ [0.5ex] 
 \hline
 Adware & eldorado, razy, gator & 1,947\\
 \hline
 Backdoor & lunam, rahack, symmi & 678\\
 \hline
 Botnet & virut, salicode, sality & 526 \\ 
 \hline
 Dropper & dunwod, gepys, doboc & 904\\
 \hline
 Ransomware & vtflooder, msil, bitman & 900\\
 \hline
 Rootkit & onjar, dqqd, shipup & 53\\
 \hline
 Spyware & mikey, qqpass, scar & 640 \\
 \hline
 Virus & nimda, shodi, hematite & 659\\
 \hline
 Total & All subtypes & 6,307\\
 \hline
\end{tabularx}
\label{testbed}
\end{table}

As our attack target, we selected three renowned  DL-based static malware detectors. All three are cited frequently by security researchers and are made available by authors through GitHub repositories.  

\begin{itemize}
    \item{\verb|MalConv|} \cite{raff2018malware}, is among the most  successful DL-based malware detectors, developed through a collaboration between the Laboratory for Physical Sciences (LPS) and NVIDIA. The model incorporates a deep convolutional neural network architecture that is trained on approximately half a million malware binaries and achieves an area under the ROC curve (AUC) of 98.5\% on an unseen test set.
    \item{\verb|NonNeg|} \cite{fleshman2019non} is a successor of MalConv developed by LPS, which modifies MalConv's architecture with non-negative weight constraints. The model was trained on 2 million malware binaries and obtained the AUC of 95.3\% on a holdout sample.
    \item{\verb|ConvNet|} \cite{krvcal2018deep} was developed by Avast research group and features a deeper neural network than MalConv and NonNeg, with a total of eight layers. It was trained on 20 million proprietary malware samples from Avast and achieved 70.4\% AUC.
\end{itemize}

Both MalConv and NonNeg were featured as recent malware detector architectures in an AME competition hosted by Endgame in 2019 \cite{anderson_machine_2019}. It is important to note that all malware samples in our dataset were recognized as malware by all three anti-malware models. Following \cite{fleshman2019non,anderson2018learning}, we adopt evasion rate as our evaluation criterion. The evasion rate of an AME method against a given anti-malware is defined as follows: 
\begin{equation}
  Evasion\ Rate = \frac{|E\cap F|}{N}
\end{equation}
where $E$ and $F$ denote the sets of evasive and functional modified malware obtained from the AME method, respectively. $N$ denotes the total number of malware samples given as input to the AME method. This statistic yields the efficacy of a given AME method in evading a malware detector. We use this metric to evaluate \malrnn against other benchmark methods later in this section.

\subsection{Experiment Setup}
\label{experiment_setup}
We conduct three different experiments. In the first experiment, we examine the number of attempts \malrnn requires to generate evasive variants. In the second experiment, we measure the changes of MalRNN's performance by varying the append size for each malware category. Finally, in the third experiment, we compare \malrnn's performance on all three malware detectors to that of other AME benchmarks for a fixed append volume (determined in our second experiment). For comparison, we identified two state-of-the-art binary black-box and one black-box AME benchmarks:
 
 \begin{itemize}
    \item{\verb|Random Append (RA)|} \cite{suciu2019exploring,castroandschmitt2019armed}: Appends sequences of random bytes to the end of a malware sample until the evasion occurs.
    \item{\verb|Benign Append (BA)|} \cite{castroandbiggio2019poster}: Appends random sections from benign files to the end of a malware sample until evasion occurs.
    \item{\verb|Enhanced Benign Append (EBA)|} \cite{chenB2019adversarial}: Appends specific byte sequences that lower the confidence score of the anti-malware in a brute-force manner.
\end{itemize}

It is worth noting that since EBA requires access to the confidence score, it is qualified as black-box and has an unfair advantage compared to the other two benchmarks and our proposed method. The following subsections describe each experiment and its corresponding results in detail.

\subsection{Can \malrnn Learn to Generate Evasive Variants?}
\begin{figure}[h]
        \includegraphics[width=0.45\textwidth]{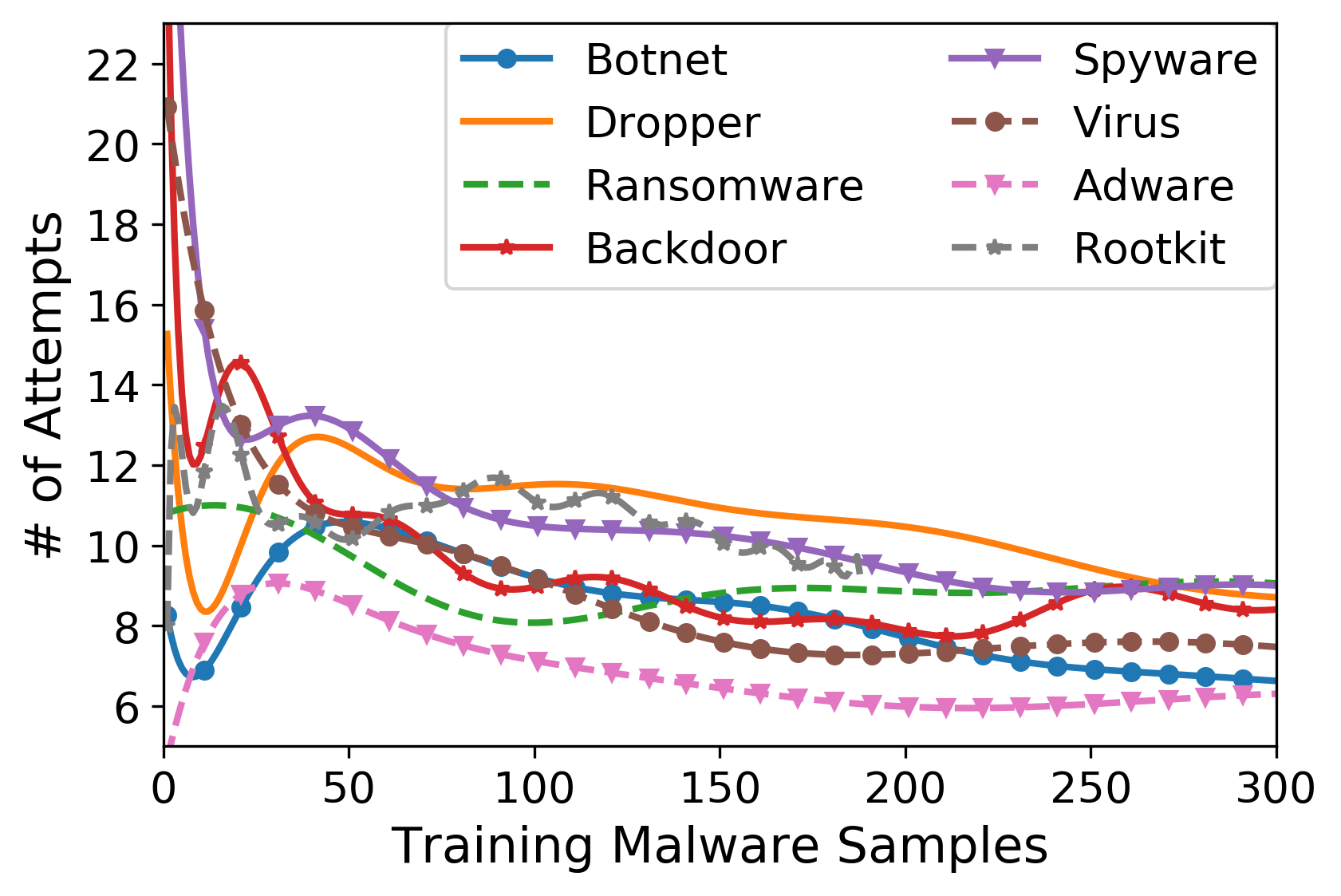}
        \caption{Running average of the number of iterations required to bypass the anti-malware engine for each sample.}
    \label{training_iterations}
\end{figure}

 It is often desirable to verify if a machine learning model learns during training by monitoring the training loss or number of iterations required to solve the problem at hand. In order to assess whether \malrnn learns to generate evasive bytes, we monitor the number of attempts (i.e, iterations) required for evasion during the training of \malrnn (Figure \ref{training_iterations}).
 
 As seen in Figure \ref{training_iterations}, when training starts \malrnn needs around 20 attempts to modify a given malware sample such that it can evade the anti-malware. However, as the training proceeds, this number significantly decreases. As a result, at the latest stages of training (after processing almost 300 malware samples) the number of required attempts reduces to around eight. This behavior is consistent among all eight categories and suggests that \malrnn improves during the training process and learns to generate evasive content. 

\subsection{How Does the Append Size Affect the Evasion Rate?}
As noted, very large append sizes can defeat the purpose of developing an effective AME method that is able to accomplish an evasion attack through \textit{minimal} modification of the original malware. As such, in practice, it is crucial to limit the maximum append size of AME methods. To empirically observe the effect of append size on the evasion rate, we track the changes in evasion rate for various append sizes in the virus category as it is one of the most damaging malware types. Table \ref{append_volume} summarizes the results.  

\begin{table}[h]
\caption{Evasion rates and number of required training iterations obtained at different append sizes}
\small
\begin{tabularx}{0.5\textwidth} { 
  | >{\centering\arraybackslash}X 
  | >{\centering\arraybackslash}X 
  | >{\centering\arraybackslash}X 
  | >{\centering\arraybackslash}X 
  | >{\centering\arraybackslash}X | }
 \hline
 \textbf{AVG Append Size (\%)} & \textbf{AVG Append Size (KB)} & \textbf{\# of Evaded Samples} & \textbf{Evasion Rate} & \textbf{\# of Training Iterations} \\
 \hline
 5 & 7.5 & 763 & 82.4\% & 14,357\\
 \hline
 10 & 15 & 862 & 93.09\% & 8,588\\
 \hline
 20 & 30 & 882 & 95.25\% & 5,133\\
 \hline
 40 & 66 & 906 & 97.84\% & 3,169\\
 \hline
 80 & 132.8 & 910 & 98.27\% & 3,065\\
 \hline
 100 & 166 & 912 & 98.49\% & 3,205\\
 \hline
 120 & 199.2 & 919 & 99.24\% & 2,840\\
 \hline
 180 & 298.8 & 921 & 99.46\% & 2,406\\
 \hline
\end{tabularx}
\label{append_volume}
\end{table}

\begin{table*}[t]

\small
\caption{Comparing \malrnn's performance on three renowned DL-based anti-malware detectors with black-box AME benchmark methods across eight malware categories}
\begin{center}
\begin{tabularx}{1\textwidth} {
  | >{\centering\arraybackslash}p{1.1cm} 
  | >{\centering\arraybackslash}p{1.1cm} 
  | >{\centering\arraybackslash}X
  | >{\centering\arraybackslash}p{1.3cm} 
  | >{\centering\arraybackslash}p{0.9cm} 
  | >{\centering\arraybackslash}X
  | >{\centering\arraybackslash}X 
  | >{\centering\arraybackslash}X
  | >{\centering\arraybackslash}X
  | >{\centering\arraybackslash}p{0.8cm} 
  | >{\centering\arraybackslash}p{0.85cm}  |}
 \hline
 \textbf{Detector} & \textbf{Method} & \textbf{Adware} & \textbf{Backdoor} & \textbf{Botnet} & \textbf{Dropper} & \textbf{Ransomware} & \textbf{Rootkit} & \textbf{Spyware} & \textbf{Virus} & \textbf{Average} \\
 \hline
 
 \multirow{3}{*}{\textbf{Malconv}} & RA & 14.34\% & 9.88\% & 8.56\% & 14.16\% & 11.78\% & 13.21\% & 10.16\% & 11.53\% & 12.29\% \\ \cline{2-11}
 & BA & 49.15\% & 41.30\% & 20.34\% & 41.92\% & 38.44\% & 11.32\% & 35.31\% & 28.22\% & 39.43\% \\ \cline{2-11}
 & EBA & \textbf{75.55\%} & 68.29\% & 46.58\% & 69.69\% & \textbf{80.22\%} & 56.60\% & 65.31\% & 61.76\% & 69.54\% \\ \cline{2-11}
 & \textbf{MalRNN} & 68.75\% & \textbf{72.72}\% & \textbf{53.66}\% & \textbf{90\%} & 64.28\% & \textbf{69.23\%} & \textbf{80\%} & \textbf{85.71\%} & \textbf{73.24\%} \\ \cline{2-11}
 \hline
 
 \multirow{3}{*}{\textbf{NonNeg}} & RA & 0.67\% & 0.44\% & 0.19\% & 1.00\% & 0.44\% & 5.66\% & 0.63\% & 0.76\% & 0.67\% \\ \cline{2-11}
 & BA & 96.61\% & 99.41\% & 99.05\% & 94.91\% & 99.00\% & 90.57\% & 93.91\% & 88.47\% & 96.04\% \\ \cline{2-11}
 & EBA & 96.10\% & 94.40\% & 95.25\% & 98.78\% & 96.56\% & \textbf{100\%} & 94.38\% & 89.38\% & 95.45\% \\ \cline{2-11}
 & \textbf{MalRNN} & \textbf{99.87\%} & \textbf{100\%} & \textbf{100\%} & \textbf{100\%} & \textbf{99.87\%} & \textbf{100\%} & \textbf{100\%} & \textbf{100\%} & \textbf{99.97\%} \\ \cline{2-11}
 \hline
 
 \multirow{3}{*}{\textbf{ConvNet}} & RA & 30.71\% & 25.96\% & 69.01\% & 26.77\% & 10.67\% & 16.98\% & 46.88\% & 54.17\% & 33.95\% \\ \cline{2-11}
 & BA & 33.23\% & 27.43\% & 66.16\% & 35.62\% & 17.67\% & 35.85\% & 47.03\% & 49.92\% & 36.64\% \\ \cline{2-11}
 & EBA & 38.46\% & 35.29\% & 43.75\% & 47.83\% & 24.00\% & 45.28\% & 46.3\% & 51.22\% & 40.03\% \\ \cline{2-11}
 & \textbf{MalRNN} & \textbf{76.49\%} & \textbf{100\%} & \textbf{87.1\%} & \textbf{69.23\%} & \textbf{35.56\%} & \textbf{64.15\%} & \textbf{73.8\%} & \textbf{70.59\%} & \textbf{72.03\%} \\ \cline{2-11}
 \hline
 
 \multirow{3}{*}{\textbf{All Three}} & RA & 0.00\% & 0.00\% & 0.00\% & 0.55\% & 0.00\% & 1.89\% & 0.00\% & 0.00\% & 1.49\% \\ \cline{2-11}
 & BA & 15.56\% & 14.31\% & 5.30\% & 5.63\% & 9.33\% & 5.66\% & 3.75\% & 0.00\% & 8.51\% \\ \cline{2-11}
 & EBA & 23.52\% & 23.15\% & 20.53\% & 19.58\% & 23.44\% & 15.09\% & 34.69\% & 22.91\% & 22.86\% \\ \cline{2-11}
 & \textbf{MalRNN} & \textbf{34.77\%} & \textbf{54.28\%} & \textbf{23.57\%} & \textbf{46.57\%} & \textbf{29.33\%} & \textbf{41.51\%} & \textbf{45.47\%} & \textbf{34.75\%} & \textbf{38.78\%} \\ 
 \hline
\end{tabularx}
\end{center}
\label{experiment_results}
\end{table*}

\begin{figure}[h]
    \centering
    \includegraphics[width=0.35\textwidth]{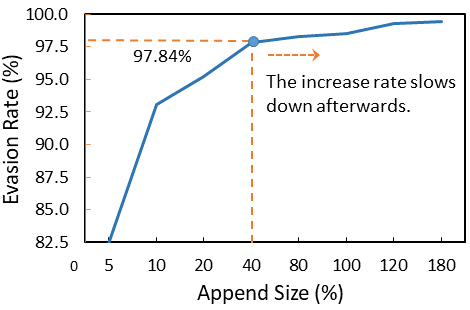}
    \caption{Evasion rate vs. append volume}
    \label{evasion_v_append}
\end{figure}

Two major observations are made from Table \ref{append_volume}. First, it is seen that by appending only 7.5 KB on average to an original malware binary, MalRNN is able to achieve the evasion rate of 82.4\%. This speaks to the effectiveness of the bytes generated by the proposed method, as will be thoroughly investigated in our third experiment. Second, and more importantly, as the append size increases from 5\% to 40\% of the original malware size, the evasion rate rapidly increases to 97.84\%. After this point, the rate of increase almost stabilizes. Also, the total number of training iterations required to evade the anti-malware decreases and exhibits the same behavior at 40\% append size. Figure \ref{evasion_v_append} visualizes this behavior by plotting the evasion rate against the changes in append size.

The 40\% append size has been shown as an elbow point, which denotes where the evasion rate stops to increase significantly. We thus fix the append size for all methods in the benchmark evaluations to 40\%. Although our proposed model yields satisfactory results at much lower append sizes (i.e., 5\% and 10\%), we selected 40\% append size in favor of the benchmark methods involved in our third experiment. Moreover, even though our model implements the black-box threat model, the amount of bytes it appends are comparable to white-box gradient-based attacks in \cite{kolosnjaji2018adversarial}, which is around 1\% to achieve 60-70\% evasion rate.

\subsection{How Does MalRNN Compare to the State-of-the-Art Black-Box AME Benchmarks?}
    
We conduct four benchmark evaluations, each focusing on specific malware detectors. The first three evaluations compare \malrnn's ability to conduct targeted AME attacks on a specific DL-based malware detector (i.e., MalConv, NoNeg, or ConvNet) individually. The last benchmark evaluation targets \malrnn's capability to evade \textit{all three} anti-malware engines simultaneously. That is, the evasion occurs only if the variant can successfully evade all three malware detectors. Such a benchmark evaluation allows us to verify \malrnn's generalizability to different DL-based models. To provide evasion rates specific to each category, we conducted benchmark evaluations separately on each malware category. Table \ref{experiment_results} summarizes the results of all four benchmark evaluations. 

\begin{figure}[h]
    \centering
    \includegraphics[width=0.45\textwidth]{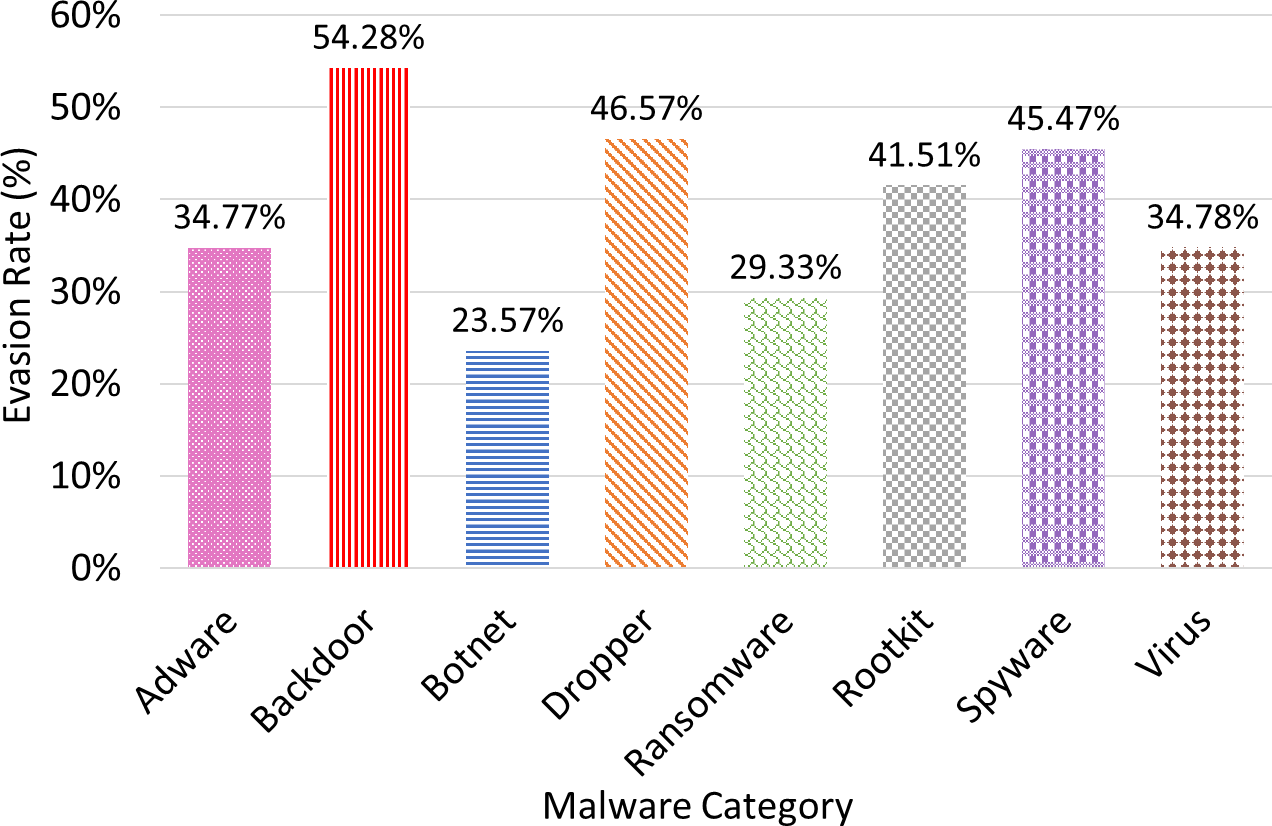}
    \caption{\malrnn's averaged evasion rate against three DL-based malware detectors across eight malware types}
    \label{experiment_graph}
\end{figure}

From Table \ref{experiment_results}, it is observed that \malrnn outperforms all other AME benchmarks in almost all of the categories for all three malware detectors. Interestingly, not only does MalrNN outperform its binary black-box AME counterparts, but it also outperforms EBA, which has access to the confidence scores, with the exception of adware and ransomware for evading MalConv. In addition to comparison with AME benchmarks, it is helpful to measure the performance of \malrnn on evading all three DL-based malware detectors across all eight malware types. Figure \ref{experiment_graph} illustrates our \malrnn's evasion rate for collectively evading all three malware detectors for each malware type.

As shown in Figure \ref{experiment_graph}, ransomware and botnet have the lowest overall evasion rate with 29.33\% and 23.57\%, respectively, which may suggest that these categories are less sensitive to AME append attacks. This could be attributed to the fact that ransomware binaries have significant sections dedicated to data encryption routines, which could be uniquely distinguished with DL-based classifiers. Similarly, botnet binaries are often unique in the sense that they incorporate a considerable amount of code devoted to establishing and maintaining the network of malicious devices on the internet. Such unique characteristics can render adversarial modifications less effective in causing these types of malware to evade. On the contrary, it is also observed that backdoor and dropper with 54.28\% and 46.57\%, respectively, have the highest evasion rate. This suggests that, overall, DL-based anti-malware models may be more susceptible to modifications of backdoor and dropper samples. This aligns with the fact that backdoor samples often contain malicious binary that is embedded into a variety of benign programs to bypass regular authentication and provide remote unauthorized access to a system. As a result, their content may be similar to non-detrimental content that are more likely to evade the DL-based anti-malware models. Similarly, droppers are also benign-looking malicious tools that are designed to embed other hidden malicious code (e.g., virus) to bypass anti-malware engines. Consequently, both backdoor and dropper malware types are difficult to identify for DL-based anti-malware models that operate on the entire binary content with large portions of benign code. This finding aligns with the intuition that crafting adversarial examples for malware executables that are already embedded in benign executables could be less difficult than other malware categories with larger portions of conspicuously malicious content (e.g., botnet and ransomware).

\section{Conclusion and Future Work}
Recently, static DL-based malware detectors have shown promise in detecting unseen malware without manual rule definition and feature engineering. However, they can themselves be vulnerable to AME attacks. We can strengthen these anti-malware engines by emulating AME attacks. Automating this process is crucial for improving anti-malware engines at a higher pace. Current approaches to this end unrealistically assume full or partial knowledge about targeted anti-malware. In this study, by treating adversarial malware generation as language modeling, we developed a novel method, MalRNN, to craft adversarial examples without requiring any knowledge of the targeted anti-malware. MalRNN directly learns a language model on binary executables and generates effective benign-looking byte sequences that can evade several DL-based anti-malware models simultaneously. MalRNN neither depends on the gradients of a differentiable anti-malware model, nor on the confidence score received from the anti-malware engine. The results signify the vulnerability of DL-based anti-malware models to adversarial append attacks and reveal that significant future research in this area is needed. Future research is needed for devising more sophisticated AME methods. One promising direction is extending the perturbations from append attacks to editing modifications to help provide more powerful AME methods. However, it should be noted that it is often harder to ensure the functionality of the malware variants obtained from editing modifications as opposed to additive ones. 

Due to nature of our study, its dual use is crucial to attend to. MalRNN contributes to emulating adversarial attacks as a viable defense mechanism to gain insight on the adversary’s capabilities. Though the ultimate goal of our study is reinforcing the robustness of anti-malware engines, precautionary measures should be taken to monitor and prevent large scale misuse of such AI techniques during the deployment of technology. Software-as-service deployment is one way to provide the monitoring so that the benevolent usage of the technology outweighs its malicious usage.

\section{ Acknowledgments}
We would like to thank Hyrum Anderson from Microsoft for valuable discussions and feedback. We also thank VirusTotal for providing the malware dataset and granting access to the APIs for functionality assessment.
This material is based upon work supported by the National Science Foundation (NSF) under the grants SaTC-1936370, CICI-1917117, and SFS-1921485.

\bibliography{bibliography}

\begin{thebibliography}{}

\bibitem[\protect\citeauthoryear{Anderson \bgroup et al\mbox.\egroup
  }{2018}]{anderson2018learning}
Anderson, H.~S.; Kharkar, A.; Filar, B.; Evans, D.; and Roth, P.
\newblock 2018.
\newblock Learning to evade static pe machine learning malware models via
  reinforcement learning.
\newblock {\em arXiv preprint arXiv:1801.08917}.

\bibitem[\protect\citeauthoryear{Anderson}{2019}]{anderson_machine_2019}
Anderson, H.~S.
\newblock 2019.
\newblock Machine {Learning} {Static} {Evasion} {Competition}.

\bibitem[\protect\citeauthoryear{Awad, Nassar, and
  Safa}{2018}]{awad2018modeling}
Awad, Y.; Nassar, M.; and Safa, H.
\newblock 2018.
\newblock Modeling malware as a language.
\newblock In {\em 2018 IEEE International Conference on Communications (ICC)},
  1--6.
\newblock IEEE.

\bibitem[\protect\citeauthoryear{Belletti, Chen, and
  Chi}{2019}]{belletti2019quantifying}
Belletti, F.; Chen, M.; and Chi, E.~H.
\newblock 2019.
\newblock Quantifying long range dependence in language and user behavior to
  improve rnns.
\newblock In {\em Proceedings of the 25th ACM SIGKDD International Conference
  on Knowledge Discovery \& Data Mining},  1317--1327.

\bibitem[\protect\citeauthoryear{Bissell and Ponemon}{2018}]{bissell2019}
Bissell, K., and Ponemon, L.
\newblock 2018.
\newblock Annual cost of cybercrime study: Unlocking the value of improved
  cybersecurity protection.

\bibitem[\protect\citeauthoryear{Castro, Biggio, and
  Dreo~Rodosek}{2019}]{castroandbiggio2019poster}
Castro, R.~L.; Biggio, B.; and Dreo~Rodosek, G.
\newblock 2019.
\newblock Poster: Attacking malware classifiers by crafting gradient-attacks
  that preserve functionality.
\newblock In {\em Proceedings of the 2019 ACM SIGSAC Conference on Computer and
  Communications Security},  2565--2567.

\bibitem[\protect\citeauthoryear{Castro, Schmitt, and
  Rodosek}{2019}]{castroandschmitt2019armed}
Castro, R.~L.; Schmitt, C.; and Rodosek, G.~D.
\newblock 2019.
\newblock Armed: How automatic malware modifications can evade static
  detection?
\newblock In {\em 2019 5th International Conference on Information Management
  (ICIM)},  20--27.
\newblock IEEE.

\bibitem[\protect\citeauthoryear{Chen \bgroup et al\mbox.\egroup
  }{2017}]{chen2017zoo}
Chen, P.-Y.; Zhang, H.; Sharma, Y.; Yi, J.; and Hsieh, C.-J.
\newblock 2017.
\newblock Zoo: Zeroth order optimization based black-box attacks to deep neural
  networks without training substitute models.
\newblock In {\em Proceedings of the 10th ACM Workshop on Artificial
  Intelligence and Security},  15--26.

\bibitem[\protect\citeauthoryear{Chen \bgroup et al\mbox.\egroup
  }{2019a}]{chenB2019adversarial}
Chen, B.; Ren, Z.; Yu, C.; Hussain, I.; and Liu, J.
\newblock 2019a.
\newblock Adversarial examples for cnn-based malware detectors.
\newblock {\em IEEE Access} 7:54360--54371.

\bibitem[\protect\citeauthoryear{Chen \bgroup et al\mbox.\egroup
  }{2019b}]{chenY2019training}
Chen, Y.; Wang, S.; She, D.; and Jana, S.
\newblock 2019b.
\newblock On training robust pdf malware classifiers.
\newblock {\em arXiv preprint arXiv:1904.03542}.

\bibitem[\protect\citeauthoryear{Demetrio \bgroup et al\mbox.\egroup
  }{2019}]{demetrio2019explaining}
Demetrio, L.; Biggio, B.; Lagorio, G.; Roli, F.; and Armando, A.
\newblock 2019.
\newblock Explaining vulnerabilities of deep learning to adversarial malware
  binaries.
\newblock {\em arXiv preprint arXiv:1901.03583}.

\bibitem[\protect\citeauthoryear{Demetrio \bgroup et al\mbox.\egroup
  }{2020}]{demetrio2020efficient}
Demetrio, L.; Biggio, B.; Lagorio, G.; Roli, F.; and Armando, A.
\newblock 2020.
\newblock Efficient black-box optimization of adversarial windows malware with
  constrained manipulations.
\newblock {\em arXiv preprint arXiv:2003.13526}.

\bibitem[\protect\citeauthoryear{Dey and Salemt}{2017}]{dey2017gate}
Dey, R., and Salemt, F.~M.
\newblock 2017.
\newblock Gate-variants of gated recurrent unit (gru) neural networks.
\newblock In {\em 2017 IEEE 60th international midwest symposium on circuits
  and systems (MWSCAS)},  1597--1600.
\newblock IEEE.

\bibitem[\protect\citeauthoryear{Dey \bgroup et al\mbox.\egroup
  }{2019}]{dey2019evadepdf}
Dey, S.; Kumar, A.; Sawarkar, M.; Singh, P.~K.; and Nandi, S.
\newblock 2019.
\newblock Evadepdf: Towards evading machine learning based pdf malware
  classifiers.
\newblock In {\em International Conference on Security \& Privacy},  140--150.
\newblock Springer.

\bibitem[\protect\citeauthoryear{Fang \bgroup et al\mbox.\egroup
  }{2019}]{fang2019evading}
Fang, Z.; Wang, J.; Li, B.; Wu, S.; Zhou, Y.; and Huang, H.
\newblock 2019.
\newblock Evading anti-malware engines with deep reinforcement learning.
\newblock {\em IEEE Access} 7:48867--48879.

\bibitem[\protect\citeauthoryear{Fleshman \bgroup et al\mbox.\egroup
  }{2019}]{fleshman2019non}
Fleshman, W.; Raff, E.; Sylvester, J.; Forsyth, S.; and McLean, M.
\newblock 2019.
\newblock Non-negative networks against adversarial attacks.
\newblock {\em arXiv preprint arXiv:1806.06108}.

\bibitem[\protect\citeauthoryear{Goldberg}{2017}]{goldberg_neural_2017}
Goldberg, Y.
\newblock 2017.
\newblock {\em Neural {Network} {Methods} for {Natural} {Language}
  {Processing}}, volume~10.

\bibitem[\protect\citeauthoryear{Goodfellow, McDaniel, and
  Papernot}{2018}]{goodfellow2018making}
Goodfellow, I.; McDaniel, P.; and Papernot, N.
\newblock 2018.
\newblock Making machine learning robust against adversarial inputs.
\newblock {\em Communications of the ACM} 61(7):56--66.

\bibitem[\protect\citeauthoryear{Hu and Tan}{2018}]{hu2018black}
Hu, W., and Tan, Y.
\newblock 2018.
\newblock Black-box attacks against rnn based malware detection algorithms.
\newblock In {\em Workshops at the Thirty-Second AAAI Conference on Artificial
  Intelligence}.

\bibitem[\protect\citeauthoryear{Irie \bgroup et al\mbox.\egroup
  }{2019}]{irie2019choice}
Irie, K.; Prabhavalkar, R.; Kannan, A.; Bruguier, A.; Rybach, D.; and Nguyen,
  P.
\newblock 2019.
\newblock On the choice of modeling unit for sequence-to-sequence speech
  recognition.
\newblock {\em Proc. Interspeech 2019}  3800--3804.

\bibitem[\protect\citeauthoryear{Kim \bgroup et al\mbox.\egroup
  }{2016}]{kim2016character}
Kim, Y.; Jernite, Y.; Sontag, D.; and Rush, A.~M.
\newblock 2016.
\newblock Character-aware neural language models.
\newblock In {\em Thirtieth AAAI conference on artificial intelligence}.

\bibitem[\protect\citeauthoryear{Kolosnjaji \bgroup et al\mbox.\egroup
  }{2018}]{kolosnjaji2018adversarial}
Kolosnjaji, B.; Demontis, A.; Biggio, B.; Maiorca, D.; Giacinto, G.; Eckert,
  C.; and Roli, F.
\newblock 2018.
\newblock Adversarial malware binaries: Evading deep learning for malware
  detection in executables.
\newblock In {\em 2018 26th European Signal Processing Conference (EUSIPCO)},
  533--537.
\newblock IEEE.

\bibitem[\protect\citeauthoryear{Kreuk \bgroup et al\mbox.\egroup
  }{2018}]{kreuk2018adversarial}
Kreuk, F.; Barak, A.; Aviv-Reuven, S.; Baruch, M.; Pinkas, B.; and Keshet, J.
\newblock 2018.
\newblock Adversarial examples on discrete sequences for beating whole-binary
  malware detection.
\newblock {\em arXiv preprint arXiv:1802.04528}.

\bibitem[\protect\citeauthoryear{Krčál \bgroup et al\mbox.\egroup
  }{2018}]{krvcal2018deep}
Krčál, M.; Švec, O.; Bálek, M.; and Jašek, O.
\newblock 2018.
\newblock Deep {Convolutional} {Malware} {Classifiers} {Can} {Learn} from {Raw}
  {Executables} and {Labels} {Only}.
\newblock In {\em International {Conference} on {Learning} {Representations}
  ({ICLR})}.

\bibitem[\protect\citeauthoryear{Madry \bgroup et al\mbox.\egroup
  }{2017}]{madry2017towards}
Madry, A.; Makelov, A.; Schmidt, L.; Tsipras, D.; and Vladu, A.
\newblock 2017.
\newblock Towards deep learning models resistant to adversarial attacks.
\newblock {\em arXiv preprint arXiv:1706.06083}.

\bibitem[\protect\citeauthoryear{Mogren and
  Johansson}{2019}]{mogren2019character}
Mogren, O., and Johansson, R.
\newblock 2019.
\newblock Character-based recurrent neural networks for morphological
  relational reasoning.
\newblock {\em Journal of Language Modelling} 7(1):139--170.

\bibitem[\protect\citeauthoryear{Ono, Utiyama, and
  Sumita}{2019}]{ono-etal-2019-hybrid}
Ono, J.; Utiyama, M.; and Sumita, E.
\newblock 2019.
\newblock Hybrid data-model parallel training for sequence-to-sequence
  recurrent neural network machine translation.
\newblock In {\em Proceedings of The 8th Workshop on Patent and Scientific
  Literature Translation},  4--12.
\newblock Dublin, Ireland: European Association for Machine Translation.

\bibitem[\protect\citeauthoryear{Park, Khan, and
  Yener}{2019}]{park2019generation}
Park, D.; Khan, H.; and Yener, B.
\newblock 2019.
\newblock Generation \& evaluation of adversarial examples for malware
  obfuscation.
\newblock In {\em 2019 18th IEEE International Conference On Machine Learning
  And Applications (ICMLA)},  1283--1290.
\newblock IEEE.

\bibitem[\protect\citeauthoryear{Qiu \bgroup et al\mbox.\egroup
  }{2019}]{qiu2019review}
Qiu, S.; Liu, Q.; Zhou, S.; and Wu, C.
\newblock 2019.
\newblock Review of artificial intelligence adversarial attack and defense
  technologies.
\newblock {\em Applied Sciences} 9(5):909.

\bibitem[\protect\citeauthoryear{Raff \bgroup et al\mbox.\egroup
  }{2018}]{raff2018malware}
Raff, E.; Barker, J.; Sylvester, J.; Brandon, R.; Catanzaro, B.; and Nicholas,
  C.~K.
\newblock 2018.
\newblock Malware detection by eating a whole exe.
\newblock In {\em Workshops at the Thirty-Second AAAI Conference on Artificial
  Intelligence}.

\bibitem[\protect\citeauthoryear{Rosenberg \bgroup et al\mbox.\egroup
  }{2019}]{rosenberg2019defense}
Rosenberg, I.; Shabtai, A.; Elovici, Y.; and Rokach, L.
\newblock 2019.
\newblock Defense methods against adversarial examples for recurrent neural
  networks.
\newblock {\em arXiv preprint arXiv:1901.09963}.

\bibitem[\protect\citeauthoryear{Suciu, Coull, and
  Johns}{2019}]{suciu2019exploring}
Suciu, O.; Coull, S.~E.; and Johns, J.
\newblock 2019.
\newblock Exploring adversarial examples in malware detection.
\newblock In {\em 2019 IEEE Security and Privacy Workshops (SPW)},  8--14.
\newblock IEEE.

\bibitem[\protect\citeauthoryear{Tounsi and Rais}{2018}]{tounsi2018survey}
Tounsi, W., and Rais, H.
\newblock 2018.
\newblock A survey on technical threat intelligence in the age of sophisticated
  cyber attacks.
\newblock {\em Computers \& security} 72:212--233.

\bibitem[\protect\citeauthoryear{Zuo \bgroup et al\mbox.\egroup
  }{2018}]{zuo2018neural}
Zuo, F.; Li, X.; Young, P.; Luo, L.; Zeng, Q.; and Zhang, Z.
\newblock 2018.
\newblock Neural machine translation inspired binary code similarity comparison
  beyond function pairs.
\newblock {\em arXiv preprint arXiv:1808.04706}.

\end{thebibliography}
\bibliographystyle{aaai}

\appendix
\label{app_A}
\section{Appendix A - MalRNN Specifications}
Both encoder and decoder in MalRNN consist of 100 GRUs with Tanh activation functions as their building blocks. All biases of the GRU units were initialized to zero. The dimension of the emebedding layer (i.e., vocabulary size) in the encoder was set to 256. Also, the size of the fully connected output layer in the decoder was set to 128. Parameter settings of \malrnn were fixed throughout all experiments. The learning rate and systematic sampling rate were set to $1{e^-2}$ and $1{e^-3}$, respectively. Also, the batch size was fixed to 10 throughout all experiments. The maximum append size was set to 40\% for all benchmark methods and MalRNN. Lastly, the maximum number of attempts was set to 50 for all benchmark methods and MalRNN. To facilitate reproducability, the code, corresponding dataset, and benchmark methods were made available on GitHub at  https://github.com/johnnyzn/MalRNN.

\end{document}